\documentclass{PoS}

\title{Kaon programme at CERN: recent results}

\ShortTitle{Kaon programme at CERN: recent results}

\author{\speaker{Evgueni Goudzovski}\thanks{On behalf of the NA48 and NA62 collaborations.}\\
        School of Physics and Astronomy, University of Birmingham\\
        E-mail: \email{eg@hep.ph.bham.ac.uk}}

\abstract{Searches for lepton flavour and lepton number violation in
kaon decays by the NA48/2 and NA62 experiments at CERN are
presented. A precision measurement of the helicity-suppressed ratio
$R_K$ of the $K^\pm\to e^\pm\nu$ and $K^\pm\to\mu^\pm\nu$ decay
rates has been performed using the full data set collected by the
NA62 collaboration in 2007. The data sample amounted to 145,958
reconstructed $K^\pm\to e^\pm\nu$ candidates with 11.0\% background
contamination. The result is $R_K=(2.488\pm0.010)\times10^{-5}$,
which is in agreement with the Standard Model expectation. An
improved upper limit on the lepton number violating
$K^\pm\to\pi^\mp\mu^\pm\mu^\pm$ decay rate is also presented.}

\FullConference{The 2011 Europhysics Conference on High Energy Physics-HEP 2011,\\
        July 21-27, 2011\\
        Grenoble, Rh\^one-Alpes France}

\begin{document}

\section*{Introduction}

Decays of pseudoscalar mesons to light leptons
($P^\pm\to\ell^\pm\nu$, denoted $P_{\ell 2}$) are helicity
suppressed in the Standard Model (SM). The specific ratios of decay
rates can be computed very precisely: the SM prediction for the
ratio $R_K=\Gamma(K_{e2})/\Gamma(K_{\mu 2})$ is~\cite{ci07}
\begin{equation}
\label{Rdef} R_K^\mathrm{SM} = ({m_e}/{m_\mu})^2
\left[(m_K^2-m_e^2)/(m_K^2-m_\mu^2)\right]^2 (1 + \delta
R_{\mathrm{QED}})= (2.477 \pm 0.001)\times 10^{-5},
\end{equation}
where $\delta R_{\mathrm{QED}}=(-3.79\pm0.04)\%$ is an
electromagnetic correction. Within certain two Higgs doublet models
(2HDM type II), $R_K$ is sensitive to lepton flavour violating (LFV)
effects via the charged Higgs boson ($H^\pm$) exchange~\cite{ma08},
the dominant contribution being
\begin{equation}
R_K^\mathrm{LFV}/ R_K^\mathrm{SM}\simeq 1+({M_K}/{M_H})^4
({M_\tau}/{M_e})^2 |\Delta _R^{31}|^2\tan^6\beta, \label{rk_lfv}
\end{equation}
where $\tan\beta$ is the ratio of the two Higgs vacuum expectation
values, and $|\Delta_{R}^{31}|$ is the mixing parameter between the
superpartners of the right-handed leptons, which can reach $\sim
10^{-3}$. This can enhance $R_K$ by ${\cal O}(1\%)$ without
contradicting any known experimental constraints. A precise $R_K$
measurement based on the full data sample ($\sim 10$ times the world
sample) collected by the CERN NA62 experiment in 2007 is presented,
superseding an earlier result~\cite{la11} based on a partial data
set.

The decay $K^\pm\to\pi^\mp\mu^\pm\mu^\pm$ violating lepton number by
two units can proceed via a neutrino exchange if the neutrino is a
Majorana particle~\cite{zu00}; it has also been studied in the
context of supersymmetric models with $R$-parity
violation~\cite{li00}. The upper limit on the decay rate has been
improved using a data sample collected by the CERN NA48/2 experiment
in 2003--2004.

\section{Search for lepton flavour violation}

The measurement method is based on counting the numbers of $K_{\ell
2}$ candidates collected concurrently. The study is performed
independently for 40 data samples (10 bins of lepton momentum and 4
samples with different data taking conditions) by computing the
ratio $R_K$ as:
\begin{equation}
R_K = \frac{1}{D}\cdot \frac{N(K_{e2})-N_{\rm
B}(K_{e2})}{N(K_{\mu2}) - N_{\rm B}(K_{\mu2})}\cdot
\frac{A(K_{\mu2})}{A(K_{e2})} \cdot
\frac{f_\mu\times\epsilon(K_{\mu2})}
{f_e\times\epsilon(K_{e2})}\cdot\frac{1}{f_\mathrm{LKr}},
\label{eq:rkcomp}
\end{equation}
where $N(K_{\ell 2})$ are the numbers of selected $K_{\ell 2}$
candidates $(\ell=e,\mu)$, $N_{\rm B}(K_{\ell 2})$ are the numbers
of background events, $A(K_{\mu 2})/A(K_{e2})$ is the geometric
acceptance correction, $f_\ell$ are the efficiencies of $e$/$\mu$
identification, $\epsilon(K_{\ell 2})$ are the trigger efficiencies,
$f_\mathrm{LKr}$ is the global efficiency of the liquid krypton
(LKr) calorimeter readout (used for electron identification), and
$D$ is the downscaling factor of the $K_{\mu2}$ trigger. A Monte
Carlo simulation is used to evaluate the acceptance correction and
the geometric part of the acceptances for background processes
entering the computation of $N_B(K_{\ell 2})$. Particle
identification, trigger and readout efficiencies and certain
backgrounds are measured directly from control data samples.

A large part of the selection is common to $K_{e2}$ and $K_{\mu2}$
decays, with two principal differences. $K_{\ell 2}$ kinematic
identification is based on the reconstructed squared missing mass
assuming the track to be a positron or a muon:
$M_{\mathrm{miss}}^2(\ell) = (P_K - P_\ell)^2$, where $P_K$ and
$P_\ell$ ($\ell = e,\mu$) are the kaon and lepton 4-momenta
(Fig.~1a). A selection condition
$M_1^2<M_{\mathrm{miss}}^2(\ell)<M_2^2$ is applied; $M_{1,2}^2$ vary
across the lepton momentum bins depending on resolution. Particle
identification is based on the ratio $E/p$ of energy deposit in the
LKr calorimeter to track momentum measured by the spectrometer.
Particles with $0.95<E/p<1.1$ ($E/p<0.85$) are identified as
positrons (muons).

The largest background to the $K_{e2}$ decay is the $K_{\mu2}$ decay
with a mis-identified muon ($E/p>0.95$) via the `catastrophic'
bremsstrahlung process in the LKr. To reduce the corresponding
uncertainty, the muon mis-identification probability $P_{\mu e}$ has
been measured as a function of momentum using dedicated data
samples. The numbers of selected $K_{e2}$ and $K_{\mu 2}$ candidates
are 145,958 and $4.2817\times 10^7$ (the latter pre-scaled at
trigger level). Backgrounds in the $K_{e2}$ sample integrated over
lepton momentum are summarised in Table~\ref{tab:bkg}: they have
been estimated with Monte Carlo simulations, except for the beam
halo background measured directly with dedicated data samples. The
contributions to the result's systematic uncertainty include the
uncertainties on the backgrounds, helium purity in the spectrometer
tank, acceptance correction, alignment, particle identification and
trigger efficiency. The final result of the measurement, combined
over the 40 independent samples taking into account correlations
between the systematic errors, is
\begin{equation}
R_K = (2.488\pm 0.007_{\mathrm{stat.}}\pm
0.007_{\mathrm{syst.}})\times 10^{-5} =(2.488\pm0.010)\times
10^{-5}.
\end{equation}
This value is consistent with the Standard Model expectation, and
the achieved precision dominates the world average.

\begin{figure}[tb]
\begin{center}
\resizebox{0.5\textwidth}{!}{\includegraphics{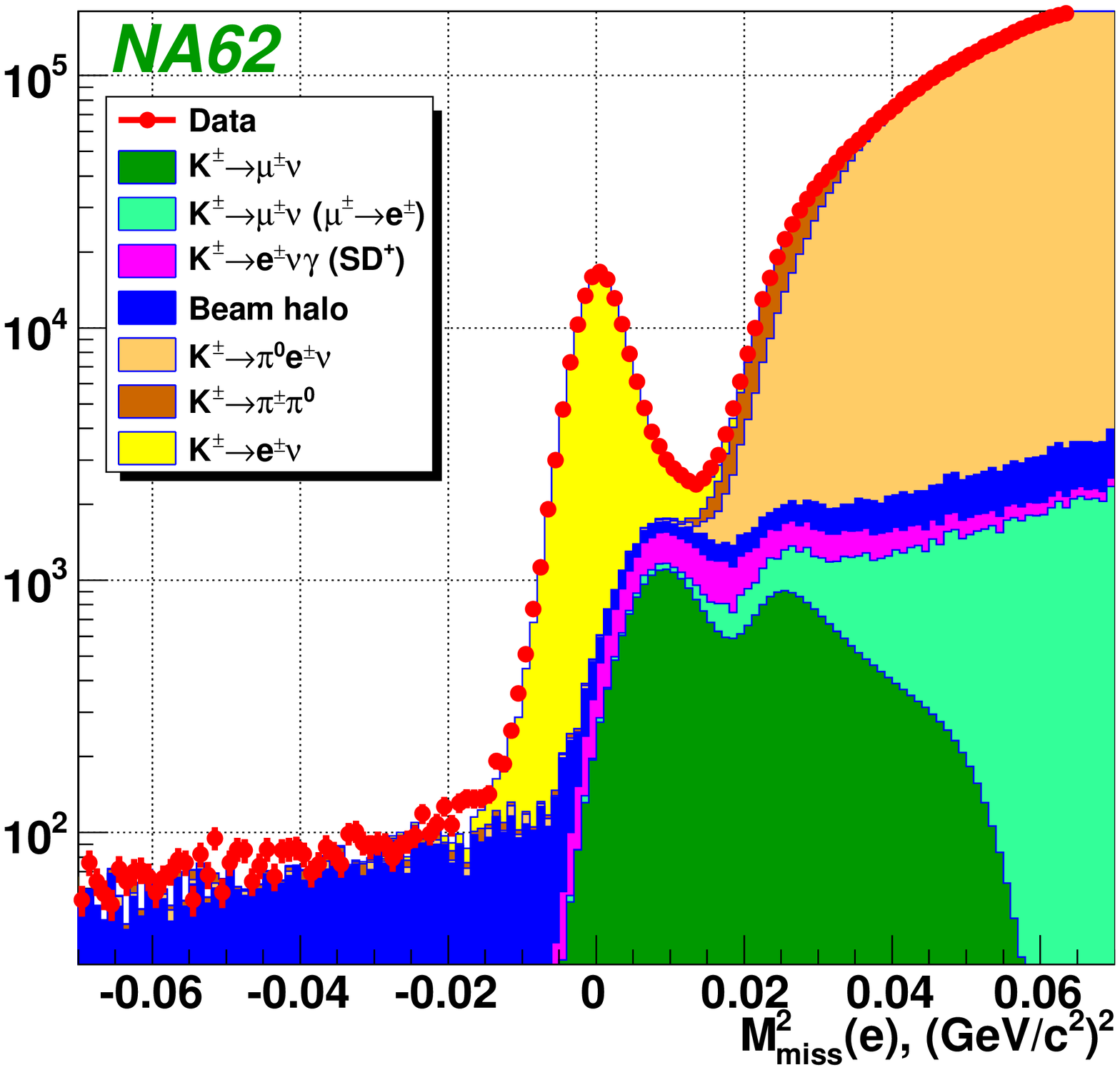}}%
{\resizebox*{0.5\textwidth}{!}{\includegraphics{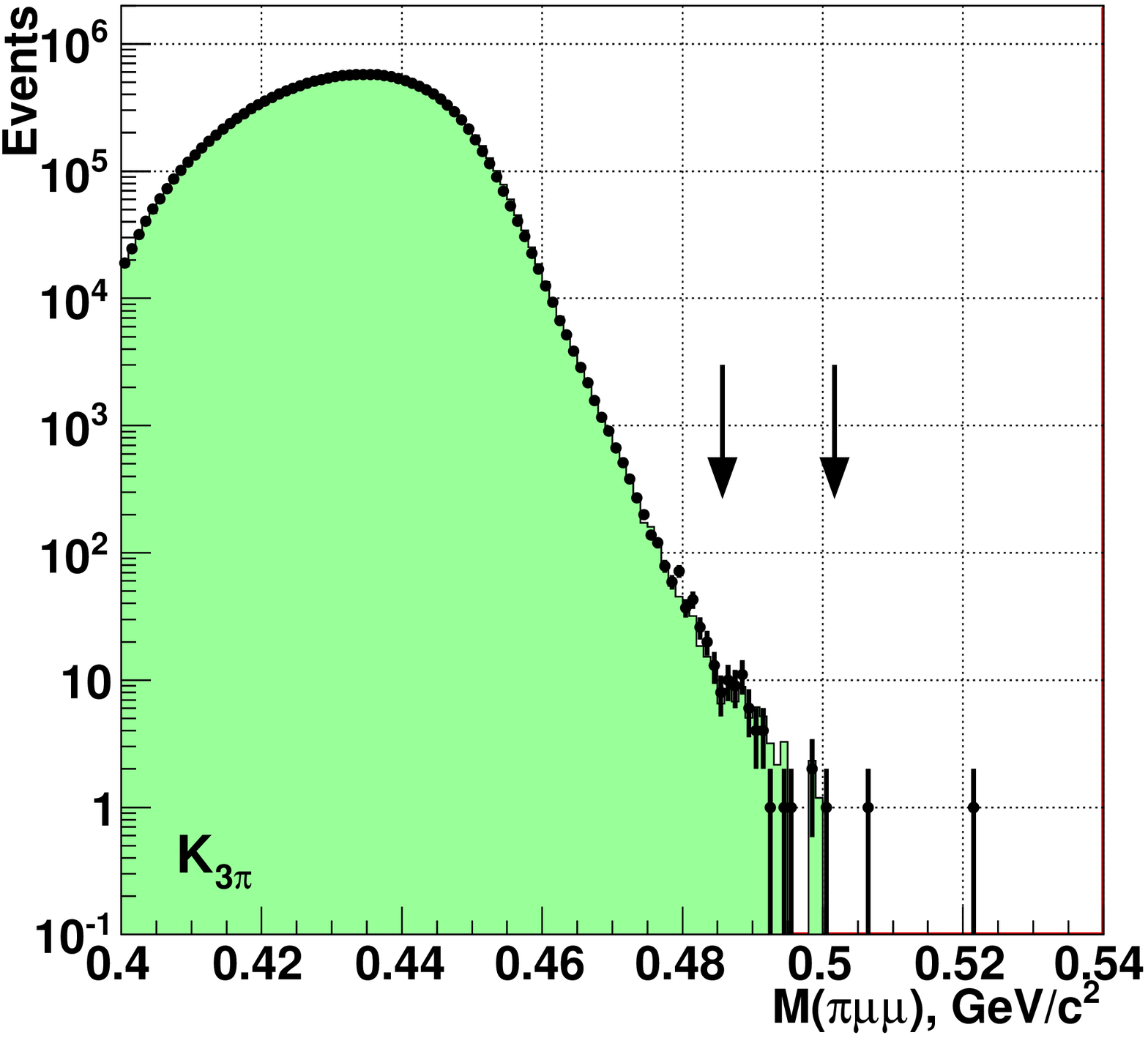}}}
\put(-240,176){\bf (a)}\put(-27,176){\bf (b)}
\end{center}
\vspace{-5mm} \caption{(a) Reconstructed squared missing mass
$M_{\mathrm{miss}}^2(e)$ distribution of the $K_{e2}$ candidates
compared with the sum of normalised estimated signal and background
components. (b) Reconstructed $M(\pi^\mp\mu^\pm\mu^\pm)$ spectra:
data (circles with error bars), $K^\pm\to 3\pi^\pm$ background
simulation (filled area).}\vspace{-2mm}
\end{figure}

\begin{table}[tb]
\begin{center}
\caption{Summary of backgrounds in the $K_{e2}$ sample.}
\label{tab:bkg} \vspace{2mm}
\begin{tabular}{l|c|l|c}
\hline Source & $N_B/N(K_{e2})$ &Source & $N_B/N(K_{e2})$\\
\hline $K_{\mu 2}$             & $(5.64\pm0.20)\%$ & $K^\pm\to\pi^\pm\pi^0$             & $(0.12\pm0.06)\%$\\
$K_{\mu 2}$ ($\mu\to e$ decay) & $(0.26\pm0.03)\%$& Beam halo                & $(2.11\pm0.09)\%$\\
$K^\pm\to e^\pm\nu\gamma~(\mathrm{SD}^+)$ & $(2.60\pm0.11)\%$ & Decays of opposite sign $K$    & $(0.04\pm0.02)\%$\\
\cline{3-4}
$K^\pm\to\pi^0 e^\pm\nu$           & $(0.18\pm0.09)\%$ & Total & $(10.95\pm0.27)\%$\\
\hline
\end{tabular}
\vspace{-6mm}
\end{center}
\end{table}

\section{Search for lepton number violation}

The $K^\pm\to\pi^\mp\mu^\pm\mu^\pm$ decay has been searched for by
reconstructing three-track vertices with no significant missing
momentum from the magnetic spectrometer information. Identification
of pion and muon candidates is performed on the basis of energy
deposition in the LKr calorimeter and the the muon detector. The
muon identification efficiency has been measured to be above $98\%$
for $p>10$~GeV/$c$, and above $99\%$ for $p>15$~GeV/$c$.

The invariant mass spectrum of the reconstructed
$\pi^\mp\mu^\pm\mu^\pm$ candidates is presented in Fig.~1b: 52
events are observed in the signal region $|M_{\pi\mu\mu}-M_K|<8~{\rm
MeV}/c^2$. The background comes from the $K_{3\pi}$ decays with
subsequent $\pi^\pm\to\mu^\pm\nu$ decay, is well reproduced by Monte
Carlo simulation, and has been estimated by the simulation to be
$(52.6\pm19.8)$ events. The quoted uncertainty is systematic due to
the limited precision of MC description of the high-mass region, and
has been estimated from the level of agreement of data and
simulation in the control mass region of (465; 485)~MeV/$c^2$. This
background estimate has been cross-checked by fitting the mass
spectrum in the region between 460 and 520~${\rm MeV}/c^2$,
excluding the signal region between 485 and 502~${\rm MeV}/c^2$,
using the maximum likelihood estimator and assuming a Poisson
probability density in each mass bin.

Conservatively assuming the expected background to be
$52.6-19.8=32.8$ events to take into account its uncertainty, the
upper limit for the possible signal is 32.2 events at 90\% CL. The
geometrical acceptance is conservatively assumed to be the smallest
of those averaged over the $K^\pm\to\pi^\pm\mu^\pm\mu^\mp$ and
$K_{3\pi}$ samples ($A_{\pi\mu\mu}=15.4\%$ and $A_{3\pi}=22.2\%$).
This leads to an upper limit of ${\rm
BR}(K^\pm\to\pi^\mp\mu^\pm\mu^\pm)<1.1\times 10^{-9}$ at 90\% CL,
which improves the best previous limit by almost a factor of 3. More
details of the analysis are presented in~\cite{ba11}.

\section{Conclusions}

The most precise measurement of lepton flavour violation parameter
$R_K$ has been performed: $R_K=(2.488\pm0.010)\times 10^{-5}$ is
consistent with the SM expectation, and can be used to constrain
multi-Higgs and fourth generation new physics scenarios. An improved
upper limit of $1.1\times 10^{-9}$ for the branching fraction of the
lepton number violating $K^\pm\to\pi^\mp\mu^\pm\mu^\pm$ decay has
been established.

\end{document}